\documentclass[12pt]{article}

\usepackage{epsfig}
\usepackage{amsfonts}
\usepackage{amsopn}
\usepackage{amsmath}
\usepackage{verbatim,amsthm}

\title
{
\vskip -50 pt
\begin{flushright}
\normalsize\rm NORDITA-2010-14
\end{flushright}
\vskip 20 pt
On the cancellation of 4-derivative terms in the Volkov-Akulov action 
}

\author{
A. A. Zheltukhin $^{a,b,c}$\thanks{e-mail: aaz@physto.se}  \\ \\
$^a$ Kharkov Institute of Physics and Technology, \\
1, Akademicheskaya St., Kharkov, 61108, Ukraine \\  \\
$^b$ Fysikum, AlbaNova, Stockholm University, \\
106 91, Stockholm, Sweden \\ \\
$^c$ NORDITA,  \\
Roslagstullsbacken 23, 106 91 Stockholm, Sweden
}

\date{}

\begin{document}

\maketitle

\begin{abstract}

Recently Kuzenko and  McCarty observed the cancellation
of 4-derivative terms in the $D=4 \  {\cal N}=1$ Volkov-Akulov 
supersymmetric action for the fermionic Nambu-Goldstone field. 
 Here is presented a simple algebraic proof of the cancellation
 based on using the Majorana bispinors and Fierz identities.
 The cancellation shows a difference between the Volkov-Akulov
action and the effective superfield action recently studied 
by Komargodski and Seiberg and containing one 4-derivative term.
We find out that the cancellation effect takes place in 
coupling of the Nambu-Goldstone fermion with the Dirac field.
Equivalence between the KS and the VA Lagrangians is proved up to the
 first order in the interaction constant of the NG fermions.

\end{abstract}

\section{Introduction}

A general approach to the construction of the phenomenological 
Lagrangians for 
the Nambu-Goldstone bosons associated with arbitrary group $G$, 
spontaneously 
broken to its subgroup $H$, was studied in the known papers 
\cite{CCWZ},\cite{V1}.
 The  Volkov's approach \cite{V1} uses the powerful Cartan's 
formalism of the exterior 
differential $\omega$-forms resulting in the invariant 
phenomenological Lagrangians of the interacting NG bosons 
\begin{equation}\label{L} 
\mathcal{L} = \frac{1}{2}Sp (G^{-1}dG)_{k}(G^{-1}dG)_{k}, \ \ \ G=KH,
\end{equation}
where the differential 1-forms $ G^{-1}dG=H^{-1}(K^{-1}dK)H + H^{-1}dH$ 
represent 
the vielbein $ (G^{-1}dG)_{k}$, and the connection $(G^{-1}dG)_{h}$ 
associated with 
the vacuum symmetry subgroup $H$.
The generalization of the NG boson conception to the fermions 
with spin $1/2$ 
associated with the spontaneous breaking of supersymmetry was 
proposed by Volkov in 
\cite{V2} and their action was consructed in \cite{VA}. 

The idea of the fermionic Nambu-Goldstone particles 
 attracts much attention and was discussed in many papers.
As the  NG fermion field  gives a nonlinear realization of supersymmetry,  its connection with the linear realization and superfields  is an important issue  within the spontaneous symmetry breaking theory. Light onto  this question  was shed in papers  \cite{Z}, \cite{IK1},  \cite{R},  \cite{LR}.  In \cite{IK1}   Ivanov and Kapustnikov  generalized  the known theorems of the nonlinear realization theory of the internal symmetries  \cite{CCWZ}  to the case of supersymmetry. They proved that any superfield could be splitted in a set of  independently transforming components with the supersymmetry parameters depending on the NG  field.  Also, they  found
that the Volkov-Akulov  Lagrangian happened to be discovered in the invariant integration measure, associated  with $x$  and $\theta$  variable  changes in the superfield action. In  \cite{IK1}  these changes were  expressed  in the form of  supersymmetry transformations, but with  their parameters substituted by  the NG  fermionic field. On the other hand in \cite{R}  Rocek derived the VA Lagrangian starting from  the scalar superfield  \cite{WZ} with the invariant constraints put on it.  As a result, he revealed the VA Lagrangian to be the component auxiliary field of the scalar superfield expressed through NG field.
In \cite{LR} Lindstrom and Rocek generalized  this approach to the case of the vector superfield  \cite{WB}.
 The connection between the linear supersymmetry and constrained superfields was  further  developed in the recent paper  by Komargodski and Seiberg \cite{Seib}, where a new superfield formalism for finding the low-energy Lagrangian of the   NG fermionic and other fields  was  proposed, and its connection  with the VA Lagrangian was considered
 \footnote{Paolo Di Vecchia attracted my attention 
to Ref. \cite{Seib}}.
The connection stimulates some questions and further studies 
in this direction.
Our interest in particular is motivated by the Kuzenko and McCarty 
paper  \cite{Kuz}, where they observed the complete cancellation
among 4-derivative terms in the $D=4 \  {\cal N}=1$ Volkov-Akulov 
supersymmetric action  \footnote{Sergei Kuzenko kindly  informed 
me about Ref. \cite{Kuz}.}.   
This cancellation shows a difference between the 
VA  \cite{VA} and KS  \cite{Seib} actions and gives rise to the 
question about the constrained  
superfield action generating an effective NG Lagrangian without
 4-derivative and higher derivative terms.
The difference between KS and VA actions originates from the different realizations of the NG fermionic field  in the VA and KS  actions. In view of  the invariance of the both actions  under supersymmetry  transformations the problem  reduces to a proper redefinition of the NG field.
  As experience shows  the finding of the explicit redefinition formula may turn out  to be  an intricate problem  due to  the presence of   higher derivative  terms of  the  NG field ( see e.g.  \cite{HK}). 
Another question is whether such a cancellation takes place in the
 NG fermion couplings with other fields.

Here we present an independent proof of the cancellation 
effect \cite{Kuz}, based on using the Majorana bispinor 
representation of the $D=4 \  {\cal N}=1$  fermionic  NG field and the 
corresponding Fierz rearrangements. We also find out that the cancellation 
effect occurs in interactions of the NG fermion with other fields. 
As a result, the 4-derivative and higher terms, associated with the
 fermionic NG 
 field, are absent in the VA  Volkov-Akulov Lagrangian with couplings \cite{VA}. 
We show that the maximal numbers of the NG fermions and their 
derivatives in the VA Lagrangian of interactions with the 
Dirac fields  equal six and three respectively. An algorithmic procedure to verify the assumption about equivalency between the KS and VA Lagrangians, based on the redefinition of the KS fermionic field, is discussed, and their equivalence up to the first order in the constant $a$, describing the interaction between the NG fermions themselves,  is proved.

In sections 2, 3, 4 we draw attention to supersymmetry 
 algebra in the Weyl 
and Majorana representations, the Volkov-Akulov action and its 
generalizations including the higher derivative terms.
In section 5 we present a new proof of the cancellations of 
4-derivative terms in the Volkov-Akulov action. 
In section 6 we find out that  the cancellation effect 
takes place in the  NG fermion couplings with the Dirac 
and other fields. The explicit formula, expressing the KS fermionic field through the VA fermion up to the first order in the interaction  constant  $a$,  is derived in section 7.

\section {Supersymmetry and superalgebra}

The  focus  here is on  the case of $D=4, \,\, {\cal N}=1$ supersymmetry which  
transformations are given by 
\begin{equation} \label{susy}
\theta'_{\alpha}= \theta_{\alpha} + \xi_{\alpha}, \ \ 
\bar\theta'_{\dot\alpha}= \bar\theta_{\dot\alpha} + \bar\xi_{\dot\alpha}, \ \
x^{'}_{\alpha\dot\alpha}=x_{\alpha\dot\alpha} + 
\frac{i}{2}( \theta_{\alpha}\bar\xi_{\dot\alpha}- \xi_{\alpha}\bar\theta_{\dot\alpha} )
\end{equation} 
 in the Weyl spinor representation 
with $x_{\alpha\dot\alpha}=x_{m}\sigma^{m}_{\alpha\dot\alpha}$
\footnote{We use algebraic agreements accepted in \cite{Z_Gra}.}. 
The Pauli matrices $\sigma_{i}$ and the identity matrix  $\sigma_{0}$ form a 
basic set  $\sigma_{m}=(\sigma_{0}, \sigma_{i})$ in the space of all $SL(2C)$ matrices.
The Lorentz covariant description uses the second  
set of the  Pauli matrices with the upper indices 
 $\tilde\sigma_{m}:=( \tilde\sigma_{0}, \tilde\sigma_{i}):=(\sigma_{0}, -\sigma_{i})$ 
\begin{equation} \label{antc}
\{\sigma_{m}, \tilde\sigma_{n}\}= -2\eta _{mn},  \ \ \ 
Sp\sigma_{m}\tilde\sigma_{n}=-2\eta_{mn},
 \ \ \  \sigma^{m}_{\alpha\dot\alpha}\tilde\sigma_{m}^{\dot\beta\beta}
=-2\delta_{\alpha}^{\beta}\delta_{\dot\alpha}^{\dot\beta}\ ,
\end{equation} 
where  $\eta_{mn}=diag(-1,1,1,1)$.
The matrices $\sigma_{m}$ and $\tilde\sigma_{m}$ are Lorentz 
invariant similarly to the tensors $\eta_{mn}$ and 
the spinor antisymmetric metric 
 $\varepsilon_{\alpha\beta}$ with the components $\varepsilon_{12}=\varepsilon^{21}=-1$.
The supersymmetry generators $Q^{\alpha}$ and their complex 
conjugate  $\bar Q^{\dot\alpha}:= -(Q^{\alpha})^{*}$ have the form 
\begin{equation} \label{gener}
Q^{\alpha}=\frac{\partial}{\partial\theta_{\alpha}} - \frac{i}{2} \bar\theta_{\dot\alpha}
\frac{\partial}{\partial x_{\alpha\dot\alpha}}, \ \ \
\bar Q^{\dot\alpha}= \frac{\partial}{\partial\bar\theta_{\dot\alpha}} 
- \frac{i}{2} \theta_{\alpha}
\frac{\partial}{\partial x_{\alpha\dot\alpha}}
\end{equation} 
and  form the supersymmetry algebra 
\begin{eqnarray}\label{susyalg}
\{ Q^{\alpha}, \bar Q^{\dot\alpha} \}=-i\frac{\partial}{\partial x_{\alpha\dot\alpha}}
=\frac{1}{2}\tilde\sigma_{m}^{\dot\alpha\alpha}P^{m}, \\
\{ Q^{\alpha},  Q^{\beta} \}=\{ \bar Q^{\dot\alpha}, \bar Q^{\dot\beta} \}
=[ Q^{\alpha},P^{m}]=[ \bar Q^{\dot\alpha},P^{m}]=0 \nonumber
\end{eqnarray}
 together with the translation generator $P^{m}=i\frac{\partial}{\partial x_{m}}$.

The supersymmetry transformations (\ref{susy}) and superalgebra (\ref{susyalg})
are presented in equivalent bispinor form after transition to the 
Majorana spinors
\begin{eqnarray}\label{gamsusy}
\delta\theta=\xi, \ \ \  \  \delta\bar\theta=\bar\xi,  \ \ \ \
\delta x_{m}=-\frac{i}{4}(\bar\xi\gamma_{m}\theta),  \ \ \ \
\{ Q_{a}, Q_{b}\}= \frac{1}{2}(\gamma_{m}C^{-1})_{ab}P^{m},
\end{eqnarray}
where  $\bar\theta=\theta^{T}C$ with the antisymmetric matrix of the 
charge conjugation $C$
\begin{eqnarray}\label{major}
C^{ab}=  \left( \begin{array}{cc}
                       \varepsilon^{\alpha\beta}&0\\
                       0 & \varepsilon_{\dot\alpha\dot\beta}
                          \end{array} \right), \ \ \ 
Q_{a}= \frac{\partial}{\partial\bar\theta^{a}} - \frac{i}{4}
 (\gamma_{m}\theta)_{a}\frac{\partial}{\partial x_{m}}.
\end{eqnarray}
The Majorana spinors and the Dirac $\gamma$-matrices 
are defined as in \cite{WB}
 \begin{eqnarray}\label{gamma}
\theta _{a}=\left(\begin{array}{c} \theta_{\alpha} \\ 
 \bar\theta^{\dot\alpha} \end{array} \right), \ \ 
\xi_{a}=\left(\begin{array}{c} \xi_{\alpha} \\ 
 \bar\xi^{\dot\alpha} \end{array} \right),  \ \  
\gamma_{m}=  \left( \begin{array}{cc}
                       0 & \sigma_{m}\\
                       \tilde\sigma_{m} & 0
                          \end{array} \right),  \ \ 
\{\gamma_{m}, \gamma_{n}\}= -2\eta _{mn}.
\end{eqnarray}

\section{The Volkov-Akulov action}

To construct the phenomenological Lagrangian of the 
Nambu-Goldstone fermions 
 the elegant formalism of the invariant 
Cartan $\omega$-forms \cite{V1},
unified with supersymmetry by Volkov, was used in \cite{VA}.
 The supersymmetry invariant differential $\omega$-forms 
in extended superspace with the Grassmannian coordinates 
$\theta_{\alpha}^{I}$ have the form 
\begin{eqnarray} \label{wforms}
\omega_{\alpha}^{I}= d\theta_{\alpha}^{I}  ,  \ \ \    
\bar\omega_{\dot\alpha I}=d\bar\theta_{\dot\alpha I}   ,  \ \ \  
\omega_{\alpha\dot\alpha}=dx_{\alpha\dot\alpha}-
\frac{i}{2}( d\theta_{\alpha}^{I} \bar\theta_{\dot\alpha I} - 
\theta_{\alpha}^{I} d\bar\theta_{\dot\alpha I}), 
\end{eqnarray}  
 where $I =1,2,...,N$ is the index of the internal $SU(N)$ symmetry.  
 
In the Majorana representation these fermionic and bosonic 
 1-forms are
\begin{eqnarray} \label{biwforms}
\omega= d\theta ,  \ \ \    
\bar\omega=d\bar\theta_  ,  \ \ \  
\omega_{m}=dx_{m}-
\frac{i}{4}( d\bar\theta\gamma_{m}\theta ).
\end{eqnarray} 
The $\omega$-forms (\ref{wforms}) were used in \cite {VA} as the building 
blocks for the construction of supersymmetric 
 actions for the interacting NG fermions. 
Posssible actions for the fermionic NG fields 
are  
constructed in the form of the wedge products of the  $\omega$-forms (\ref{wforms}),
forming  hyper-volumes imbedded in the extended superspace. The action candidates 
have to be invariant under the Lorentz and internal (unitary) symmetries. 
 In the case of the $4D$ Minkowski space the invariant action of the NG fermions 
must include the factorized volume element $d^{4}x$. This requirement restricts the 
structure of the admissible combinations of the $\omega$-forms. If 
 such a combination is given by a wedge product of the $\omega$-forms (\ref{wforms})
and their differentials, it should have the general number of the 
differentials equals four. 
The conditition is satisfied by the well known invariant  \cite{VA}
\begin{eqnarray} \label{volum}
 d^{4}V=\frac{1}{4!}\varepsilon_{mnpq}\omega^{m}\wedge\omega^{n}\wedge\omega^{p}
\wedge\omega^{q}, 
\end{eqnarray} 
where $\wedge$ is the wedge product symbol, that gives 
the  supersymmetric extension of the volume element $d^{4}x$ 
of the Minkowski space.
The supersymmetric volume (\ref{volum}) is invariant under the Lorentz
 and unitary groups. It does not contain the spinorial 
one-forms $\omega_{\alpha}^{I}$ and $\bar\omega_{\dot\alpha I}$, but they appear, 
 e.g. in  the following invariant products   \cite{VA}
\begin{eqnarray}\label{volhor}
 \Omega^{(4)}=\omega_{\alpha}^{I}\wedge\bar\omega_{\dot\beta I}\wedge 
\tilde\sigma^{\dot\beta\alpha}_{m} d\wedge\omega^{m},\ \ \ 
 {\tilde\Omega}^{(4)}=\varepsilon^{\alpha\beta}\omega_{\alpha}^{I}\wedge
\omega_{\beta}^{J}\wedge\bar\omega_{\dot\alpha I}\wedge 
\bar\omega_{\dot\beta J}\varepsilon^{\dot\alpha\dot\beta},
\end{eqnarray} 
where $d\wedge\omega^{m}$ is the exterior differential of $\omega^{m}$.
The additional important  symmetry of the invariants (\ref{volum}) 
and (\ref{volhor}) is 
their independence on the choice of the superspace coordinate realization. 
It means that the four dimensional hypersurfaces, associated with  
(\ref{volum}-\ref{volhor}), may be parametrized by various ways.
Because the Volkov's idea was to identify the Grassmannian $\theta$ 
coordinates 
with the fermionic  NG fields, associated with the spontaneous 
breaking of supersymmetry, 
they must be considered as functions of $x$. 
This requirement means transition to the non-linear realization 
of supersymmetry. 

It explains why the pullbacks of the 4-form $d^{4}V$ (\ref{volum}) or its 
generalizations (\ref{volhor}) on 
 the $4$-dimensional Minkowski subspace were proposed in \cite{VA}
to generate supersymmetric actions for the fermionic NG fields.
As a result of the observations, the differential forms  
$\omega_{m}$ (\ref{biwforms}) and  $d^{4}V$ 
are presented as 
\begin{eqnarray}\label{pullb}
\omega_{m}= 
(\delta_{m}^{n} -\frac{i}{4}\frac{\partial\bar\theta}
{\partial x_{n}}\gamma_{m}\theta)
dx_{n}=W_{m}^{n}dx_{n}, \  \  \  
d^{4}V=\det W d^{4}x.
\end{eqnarray}  
The identification of $\theta$ with the fermionic  NG field is 
 achieved by the change:
$\psi(x) =a^{-1/2}\theta(x)$, where $a$ has sense of the interaction 
 constant $[a]=L^{4}$ that introduces a supersymmetry breaking scale.
This constant restores the correct dimension $L^{-3/2}$ of the 
fermionic field $\psi(x)$ and the transition to $\psi$
in  (\ref{pullb}) and  $d^{4}V$ 
 yields the original  Volkov-Akulov action \cite{VA}
\begin{eqnarray} \label{action}
S=\frac{1}{a}\int\det W d^{4}x 
\end{eqnarray} 
with the $4\times4$ matrix $W_{m}^{n}(\psi,\partial_{m}\psi)$ defined by 
the following relations
\begin{eqnarray} \label{acterm}
W_{m}^{n}= \delta^{n}_{m} + aT_{m}^{n},\ \ \ \ 
T_{m}^{n}= -\frac{i}{4}\partial^{n}\bar\psi\gamma_{m}\psi.
\end{eqnarray}

An explicit form of the action $S$ (\ref{action}) follows from the definition 
of $\det W$
\begin{eqnarray}\label{determ}
\det W =  -\frac{1}{4!}\varepsilon_{n_{1}n_{2}n_{3}n_{4}}\varepsilon^{m_{1}m_{2}m_{3}m_{4}}
W_{m_{1}}^{n_{1}}W_{m_{2}}^{n_{2}}W_{m_{3}}^{n_{3}}W_{m_{4}}^{n_{4}},
\end{eqnarray}
where we chose $\varepsilon_{0123}=1$. 
Using  (\ref{acterm}) and (\ref{determ}) presents $S$ (\ref{action}) in the form 
\begin{eqnarray} \label{action'}
S= \int d^{4}x [\, \frac{1}{a} + T_{m}^{m} + \frac{a}{2}(T_{m}^{m}T_{n}^{n}-
T_{m}^{n}T_{n}^ {m}) + a^2 T^{(3)}+ a^3 T^{(4)}  \,],
\end{eqnarray} 
where $T^{(3)}$ and $T^{(4)}$ code the interaction terms of the NG 
fermions that are 
cubic and quartic in the fermion derivative $\partial_{m}\psi$.    
 The first term in (\ref{action'}) provides a non-zero vacuum 
expectation value 
 for the VA Lagrangian, confirming that it describes 
 the spontaneously broken supersymmetry.
In supergravity this term associates with the cosmological term \cite{VS},
\cite{Verice}.
The second term reproduces the free action for the  massless NG fermion
  $\psi(x)$ 
\begin{eqnarray} \label{dirac}
S_{0}=\int d^{4}x T_{m}^{m}
= -\frac{i}{4}\int d^{4}x\partial^{m}\bar\psi\gamma_{m}\psi,
\end{eqnarray} 
The terms $ T^{(3)}$ and  $ T^{(4)}$ cubic and respectively quartic in 
the NG fermion derivatives were  presented in \cite{VA} in the form 
\begin{eqnarray} \label{vertex}
 T^{(3)}= \frac{1}{3!}\sum_{p} (-)^{p} T_{m}^{m} T_{n}^{n} T_{l}^{l} \ ,\ \ \ \ \
T^{(4)}= \frac{1}{4!}\sum_{p} (-)^{p} T_{m}^{m} T_{n}^{n} T_{l}^{l}T_{k}^{k} \ ,
\end{eqnarray} 
where the sum  $\sum_{p}$ corresponds to the sum in all permutations of 
subindices in the products of the tensors $ T_{n}^{m}$. The terms (\ref{vertex})
 describe the vertexes with six and eight NG fermions.

\section{Higher derivative generalizations \\
of the Volkov-Akulov action}

The $\omega$-form formalism \cite{VA} yields a clear geometric 
way to extend the VA action by 
 the higher degree terms in the NG fermion derivatives.  
In general case the combinations of the $\omega$-forms (\ref{biwforms}), 
admissible for the higher order invariant actions, have to be the homogenious 
functions of the degree four in the differentials $dx$ and $d\psi$. 
The latter condition guarantees the factorization of the volume element $d^{4}x$ 
in the action integral. To restrict the number of these invariants  
the minimality condition for the degree of derivatives 
$\partial_{m}\psi$
in the general action
\begin{equation}\label{genract}
S=\int d^{4}x  L(\psi, 
\partial_{m}\psi)
\end{equation} 
 was proposed in \cite{VA}.
The minimality condition takes into account only the lowest degrees 
of the derivatives  $\partial_{m}\psi$
in the Lagrangian and corresponds to the low energy limit. 
To count the degree of $\partial_{m}\psi$ in different
invariants  observed was that these  derivatives appear from the 
differentials $d\psi$ 
in the fundamental $\omega$-forms. From this point of view there is 
an important difference among the spinor and vector 1-forms (\ref{wforms}).
The spinor one-form contain
one derivative  
$\partial_{m}\psi$,
but the vector 
form (\ref{pullb}) 
either do not contain the $\psi$ fields at all or contain one 
derivative  
$\partial_{m}\psi$ accompanied by $\psi$. 
As a result,
the whole number of the derivatives 
$\partial_{m}\psi$  with respect 
to the whole number of
 the  fermionic  NG fields  is lower in the vector differential 
one-forms than in the spinor ones.  The invariants
including the exterior differential of the $\omega$-forms 
like 
$\Omega^{(4)}$ in (\ref{volhor}) have the higher degree 
in $\partial_{m}\psi$
in comparison with the product of $\omega$-forms themselves. 
The same conclusion 
concerns the invariant  ${\tilde\Omega}^{(4)}$ including only 
the spinor forms. 

Thus, the demand of the minimality of  
the number of  the derivatives $\partial_{m}\psi$ 
in $S$ (\ref{genract}) will be  satisfied if the admissible 
invariants will contain only the vector differential 
one-forms $\omega_{m}$. 
The exact realization of the minimality condition by the
VA action fixes the latter, and solves the   
 problem of the effective action construction in the low energy limit.

\section{Cancellation of 4-derivative terms in the Volkov-Akulov action}

 For the case of  ${\cal N}=1$  supersymmetry the algebraic structure
 of the terms $T^{(3)}$ and  $T^{(4)}$ (\ref{vertex}) was analysed 
in \cite{Kuz} using  the Weyl spinor basis.
It was observed that the terms having the fourth degree 
in $\partial_{m}\psi$
and collected in  $T^{(4)}$ 
completely cancel out. 

Here we consider an alternative proof of the observation using 
the Majorana bispinor representation. In correspondence to 
representation (\ref{determ}) the term $T^{(4)}$  (\ref{vertex}) 
may be  written as  
 \begin{eqnarray}\label{8vertex}
&&T^{(4)}
= -\frac{1}{4!}\varepsilon_{n_{1}n_{2}n_{3}n_{4}}\varepsilon^{m_{1}m_{2}m_{3}m_{4}}
T_{m_{1}}^{n_{1}}T_{m_{2}}^{n_{2}}T_{m_{3}}^{n_{3}}T_{m_{4}}^{n_{4}} =
 \\
&&  -\frac{1}{4!}(\varepsilon_{n_{1}n_{2}n_{3}n_{4}}\bar\psi_{a_{1}}^{,n_{1}}
\bar\psi_{a_{2}}^{,n_{2}}\bar\psi_{a_{3}}^{,n_{3}}\bar\psi_{a_{4}}^{,n_{4}})
( \varepsilon^{m_{1}m_{2}m_{3}m_{4}}\gamma^{a_{1}b_{1}}_{m_{1}}\gamma^{a_{2}b_{2}}_{m_{2}}
\gamma^{a_{3}b_{3}}_{m_{3}}\gamma^{a_{4}b_{4}}_{m_{4}})
\nonumber \\
&& (\psi_{b_{1}}\psi_{b_{2}}\psi_{b_{3}}\psi_{b_{4}}),\nonumber
\end{eqnarray} 
where  $\gamma^{ab}_{m}=(C\gamma_{m})^{ab}$ is a symmetric matrix in 
the bispinor indices 
$(a,b=1,2,3,4)$ and the condenced 
notation $\bar\psi_{a}^{,n}:=\partial^{n}\bar\psi_{a}$ is introduced.
The product $\psi_{b_{1}}\psi_{b_{2}}\psi_{b_{3}}\psi_{b_{4}}$ in (\ref{8vertex}) 
is a completely 
antisymmetric spin-tensor of the maximal rank four since of the    
Grassmannian nature of  the spinor components $\psi_{b}$. Then  we find  
that the product  may be presented in the equivalent form as
\begin{eqnarray}\label{antisimtr}
\psi_{b_{1}}\psi_{b_{2}}\psi_{b_{3}}\psi_{b_{4}}
= - (C^{-1}_{b_{1}b_{2}}C^{-1}_{b_{3}b_{4}} + C^{-1}_{b_{1}b_{3}}C^{-1}_{b_{4}b_{2}} +
C^{-1}_{b_{1}b_{4}}C^{-1}_{b_{2}b_{3}}) \psi_{1}\psi_{2}\psi_{3}\psi_{4},
\end{eqnarray} 
where the antisymmetric matric $C^{-1}$ is inverse of the charge conjugation 
matrix $C$ (\ref{gamsusy}).
The representation (\ref{antisimtr}) collects  all spinors $\psi$ without 
derivatives in the form of a scalar multiplier. 
The  substitution of (\ref{antisimtr}) in (\ref{8vertex}) transforms it 
into  the sum of products of the bilinear spinor covariants
\begin{eqnarray}
 T^{(4)}= \frac{3}{4!} \,\Phi\,\Xi, \ \ \  \Xi:=\psi_{1}\psi_{2}\psi_{3}\psi_{4},  
\label{bilcovar}
\\
\Phi:=\varepsilon_{n_{1}n_{2}n_{3}n_{4}}\varepsilon^{m_{1}m_{2}m_{3}m_{4}}
(\bar\psi^{,n_{1}}\Sigma_{m_{1}m_{2}}\psi^{,n_{2}})
(\bar\psi^{,n_{3}}\Sigma_{m_{3}m_{4}}\psi^{,n_{4}}),\nonumber
\end{eqnarray}
where $\Sigma_{mn}:= \frac{1}{2}[\gamma_{m}, \gamma_{n}]$ are  
 the Lorentz transformation generators. 

Taking into account the well known  property of $\Sigma_{mn}$
\begin{eqnarray}\label{rotat}
\varepsilon^{m_{1}m_{2}m_{3}m_{4}}\Sigma_{m_{3}m_{4}}=-2\gamma^{5}\Sigma_{m_{1}m_{2}},
\ \ \
\gamma^{5}:=\gamma^{0}\gamma^{1}\gamma^{2}\gamma^{3}=
 \left( \begin{array}{cc}
                       -i& 0  \\
                       0 & i
                          \end{array} \right),
\end{eqnarray}
 one can  present the Lorentz invariant $\Phi$  (\ref{bilcovar}) in 
the compact form 
\begin{eqnarray}\label{bilcovar'}
\Phi=-2\varepsilon_{n_{1}n_{2}n_{3}n_{4}}
(\bar\psi^{,n_{1}}\Sigma_{m_{1}m_{2}}\psi^{,n_{2}})
(\bar\psi^{,n_{3}}\Sigma^{m_{1}m_{2}}\gamma^{5}\psi^{,n_{4}}).
\end{eqnarray}
Using representation (\ref{bilcovar'}) we shall prove the vanishing of $\Phi$. 
To this end let us recall the known Fierz relation for 
the Grassmannian spinors $\chi_{i}$
\begin{eqnarray}\label{fierz1} 
(\bar\chi_{1}\chi_{2})(\bar\chi_{3}\chi_{4})=- \frac{1}{4} \sum_{N=1}^{16}
(\bar\chi_{1}\Gamma^{A}\chi_{4})(\bar\chi_{3}\Gamma_{A}\chi_{2}),
\end{eqnarray}
where the 16 Dirac matrices $\Gamma^{A}$ and their inverse $\Gamma_{A}=(\Gamma_{A})^{-1}$, 
defined as 
\begin{eqnarray}
\Gamma^{A}:=(1,\, \gamma^{m}, \, \Sigma^{mn},  \, \gamma^{5},\,
\gamma^{5} \gamma^{m}), \label{basis} \\ 
\Gamma_{A}:=(\Gamma^{A})^{-1}=(1,\, -\gamma_{m}, 
\, -\Sigma_{mn},  \, -\gamma^{5},\, -\gamma^{5} \gamma_{m}), \nonumber
\end{eqnarray}
form the complete basis in the space of $4\times4$ matrices. 
As a result, we obtain
 \begin{eqnarray}\label{firtz2} 
\Phi=\frac{1}{2}\varepsilon_{n_{1}n_{2}n_{3}n_{4}}
 \sum_{A=1}^{16}
(\bar\psi^{,n_{1}}\Sigma_{m_{1}m_{2}}\Gamma^{A}\Sigma^{m_{1}m_{2}}\gamma^{5}\psi^{,n_{4}})
(\bar\psi^{,n_{3}}\Gamma_{A}\psi^{,n_{2}}).
\end{eqnarray}
The r.h.s of (\ref{firtz2}) includes the products of two bilinear covariants. 
The second  (right) of them  
 $(\bar\psi^{,n_{3}}\Gamma_{A}\psi^{,n_{2}})$ is either symmetric 
or antisymmetric under the permutation $n_{3}\leftrightarrow n_{2}$. 
Only the antisymmetric covariants generated by
 $\Gamma_{A}= (- \gamma_{r}, -\Sigma_{rs})$ give non-zero 
contribution to (\ref{firtz2}). 
The first (left) covariant in (\ref{firtz2}), corresponding to the above 
choice of $\Gamma_{A}$, includes either the matrix  $L_{v}$  or $L_{t}$ 
given by the expressions
\begin{eqnarray}\label{tvcovr} 
L_{v}=\Sigma_{m_{1}m_{2}}\gamma^{r}\Sigma^{m_{1}m_{2}}\gamma^{5}, \ \ \
L_{t}=\Sigma_{m_{1}m_{2}}\Sigma^{rs}\Sigma^{m_{1}m_{2}}\gamma^{5}. 
\end{eqnarray}
Using the representation of  $\Sigma_{m_{1}m_{2}}$ in the form
 $\Sigma_{m_{1}m_{2}}= (\eta_{m_{1}m_{2}}+ 
\gamma_{m_{1}}\gamma_{m_{2}})$ 
we  obtain the following relations 
 \begin{eqnarray}
\Sigma_{m_{1}m_{2}}\Gamma^{A}\Sigma^{m_{1}m_{2}}= 
4\Gamma^{A} - \gamma_{m_{1}}\gamma_{m_{2}}\Gamma^{A}\gamma^{m_{2}}\gamma^{m_{1}}, 
\label{idents} 
\\ 
\gamma_{m}\gamma^{r}\gamma^{m}= 2\gamma^{r},
\ \ \ \ 
\gamma_{m}\Sigma^{rs}\gamma^{m}=0  \nonumber
\end{eqnarray}
which show that 
\begin{eqnarray}\label{lvzero} 
L_{v}=0,  \ \ \  L_{t}= 4\Sigma^{rs}\gamma^{5}. 
\end{eqnarray}
Using the results (\ref{lvzero}) permits to present  
 (\ref{firtz2}) in  the next form
\begin{eqnarray}\label{tvcovr'} 
\Phi=-2\varepsilon_{n_{1}n_{2}n_{3}n_{4}}(\bar\psi^{,n_{1}}\Sigma^{rs}\gamma^{5}\psi^{,n_{4}})
(\bar\psi^{,n_{3}}\Sigma_{rs}\psi^{,n_{2}}).
\end{eqnarray}
Taking into account the symmetry property
$(C\Sigma^{rs}\gamma^{5})^{ab}=(C\Sigma^{rs}\gamma^{5})^{ba}$ 
and changing the summation 
indices $n_{3}\leftrightarrow n_{1}$ one can present the expression (\ref{tvcovr'})
 in the form 
\begin{eqnarray}\label{tvcovrfin}
\Phi=2\varepsilon_{n_{1}n_{2}n_{3}n_{4}}(\bar\psi^{,n_{1}}\Sigma_{rs}\psi^{,n_{2}})
(\bar\psi^{,n_{3}}\Sigma^{rs}\gamma^{5}\psi^{,n_{4}}).
\end{eqnarray}
The matching (\ref{bilcovar'}) and (\ref{tvcovrfin}) yields the expected  result
\begin{eqnarray}\label{final}
\Phi=-\Phi \ \ \ \Rightarrow  \ \ \   \Phi=0, \ \ \ \  T^{(4)}=0
\end{eqnarray}
which proves that the 4-derivative term $T^{(4)}$ (\ref{8vertex}) actually vanishes 
in agreement with the observation \cite{Kuz}.

Thus, the maximal number of derivatives present in the Volkov-Akulov 
action reduces to  three and the action takes the following form
\begin{eqnarray} \label{finaction'}
S= \int d^{4}x [\, \frac{1}{a} + T_{m}^{m} + \frac{a}{2}(T_{m}^{m}T_{n}^{n}-
T_{m}^{n}T_{n}^ {m}) + 
  \frac{a^2}{3!}\sum_{p} (-)^{p} T_{m}^{m} T_{n}^{n} T_{l}^{l} \,]
\end{eqnarray}
with  the  maximal number of NG fermions in the vertices equal six.

Matching the Lagrangian (\ref{finaction'}) and the Komargodski and Seiberg 
Lagrangian \cite{Seib}, having the form 
\begin{eqnarray} \label{KS}
{\cal L}_{KS}= -f^2 + i\partial_{\mu}\bar G\tilde\sigma^{\mu} G + 
\frac{1}{4f^2} {\bar G}^2\partial^{2} G^{2} -
\frac{1}{16f^6}  G^2{\bar G}^2\partial^{2} G^{2}\partial^{2} {\bar G}^{2},
\end{eqnarray}
shows their difference, because of the presence of one  4-derivative term
including eight NG fermions in (\ref{KS}). We shall explain that  the difference  originates from various realizations of the NG field used in the VA and KS Lagrangians.
The second question  concerns a possibility of such type cancellations in the 
NG fermion couplings with other fields.

\section{Coupling of the fermionic Nambu-Goldstone fields with the Dirac field}

Here we show that the above discussed cancellation 
of the 4-derivative terms 
also occurs in the NG fermion couplings with other fields.
It is easy to see by the application of the general Volkov's 
method \cite{V1}
 in the construction of the phenomenological Lagrangian, describing 
the NG particles interacting with other fields. 
The extension of this  method, aimed at including the 
supersymmetric couplings, 
 is based on joining of the differential $d\chi$ of a  given 
field $\chi$, carrying arbitrary  spinor 
and unitary indices, to the set of the supersymmetric $\omega$-forms 
 \cite{VA}. 
Then the above described procedure for the minimal VA action 
construction, using only the $\omega$-forms  (\ref{wforms}),
may be applied to the enlarged set of these supersymmetric one-forms. 
The only restriction on the admissible  $\chi$-terms 
 is the demand of their invariance under the Lorentz and 
the internal symmetry groups.  
 The effective actions must be the homogenious 
functions of the degree four in the 
differentials $dx, \, d\psi$ and $d\chi$, 
and generally it has to restict the number of the derivatives  
$\partial_{m}\psi$ to be less than four. 
Then the considered cancellations are not relevant.  
However, if $d\chi$ is absent in the couplings then the 
4-derivative cancellation may take place and will reduce the  
derivative $\partial_{m}\psi$ number in the coresponding vertices.

An instructive example of the described possibility gives 
the ${\cal N}=1$ minimal supersymmetric coupling of the
 fermionic NG particle with the massive Dirac field $\chi$ 
in the low energy limit \cite{VA}
\begin{eqnarray} \label{fermi1}
S= \int [ \, 
\frac{i}{2}\varepsilon_{mnpq}
(\bar\chi{\gamma^m}d\chi-d\bar\chi{\gamma^m}\chi)
\wedge\omega^{n}\wedge\omega^{p}\wedge\omega^{q} + \\
 m{\bar\chi}\chi
\varepsilon_{mnpq}\omega^{m}\wedge\omega^{n}
\wedge\omega^{p}\wedge\omega^{q} \, ].  \nonumber 
\end{eqnarray} 
The kinetic term of the Dirac field in (\ref{fermi1})
 includes the differential  $d\chi$ and the cancellation 
is absent here. The maximal number of the NG fermions at this 
term $n_{NGf}$ equals six and the maximal number $n_{NGd}$ of 
their derivatives equals three, just as in the case of the VA Lagrangian  
(\ref{finaction'}) after 4-derivative cancellation.
The mass term in (\ref{fermi1}) does not include $d\chi$ and respectively it 
includes the supervolume form $d^{4}V$ (\ref{volum}), because of the 
minimality condition. 
 Then the cancellation 
effect does work and  results in the same maximal numbers 
 $n_{NGf}=6$ and $n_{NGd}=3$ as in the kinetic term.
To present (\ref{fermi1}) in the standard notations \cite{VA}
 we substitute 
the $\omega$-forms (\ref{pullb}) in (\ref{fermi1}) and obtain
\begin{eqnarray} 
S= \int d^{4}x [\, R^{m}_{m} +a( R^{m}_{m} T^{n}_{n} - R^{m}_{n} T^{n}_{m}) + 
\frac{a^2}{2}\sum_{p} (-)^{p} R_{m}^{m} T_{n}^{n} T_{l}^{l} + \label{fermi2}
\\
\frac{a^3}{3!}\sum_{p} (-)^{p} R_{m}^{m} T_{n}^{n} T_{l}^{l}T_{k}^{k} + 
 m\bar\chi\chi\det W  \,], \nonumber
\end{eqnarray} 
where
$ R^{m}_{n}:= \frac{i}{2}( \bar\chi{\gamma^m}\partial_{n}\chi -
\partial_{n}\bar\chi{\gamma^m}\chi)$ is the kinetic term for $\chi$. 
Using the expression for $\det W$ from (\ref{finaction'}), 
 the mass term in (\ref{fermi2}) is presented as 
\begin{eqnarray} 
m\bar\chi\chi\det W = m\bar\chi\chi +
 a m\bar\chi\chi [\, T_{m}^{m} + \frac{a}{2}(T_{m}^{m}T_{n}^{n}-
T_{m}^{n}T_{n}^ {m}) + \label{fmass}  
\\
  \frac{a^2}{3!}\sum_{p} (-)^{p} T_{m}^{m} T_{n}^{n} T_{l}^{l}\, ],   
\nonumber
\end{eqnarray}
 where  $T_{m}^{n}= -\frac{i}{4}\partial^{n}\bar\psi\gamma_{m}\psi$ 
in accordance with the definition (\ref{acterm}). 

The mass term  (\ref{fmass}) contains the maximal number of 
the NG fermions 
 $n_{NGf}=6$  and respectively the derivative number $n_{NGd}=3$,
as  a consequence of the  cancellation of 4-derivative terms.
These maximal numbers  $n_{NGf}=6$ and  $n_{NGd}=3$, 
characterizing the  
structure of the interaction action (\ref{fermi1}), are the same as 
for the VA action (\ref{finaction'}).
The considered example shows that the cancellation effect takes place 
in the supersymmetric couplings containing the supervolume (\ref{volum}). 
So, we obtain that an enough condition for the 4-derivative 
cancellation in the couplings of the fermionic NG particles is 
the presence of  $d^{4}V$ (\ref{volum}) there.  
The observation sets issue on the
restoration of a  constrained superfield action with couplings which coincides with the effective VA action.

\section{Relation between the KS and \\
the VA  Lagrangians}

Despite the difference between the VA and the KS Lagrangians it seems  that they are equivalent up to the NG field redefinition.  Here we outline a straightforward way to check this assumption, and prove equivalence these Lagrangians  up to the first order in the constant $a$. 
The proof  is analogous with the one considered in \cite{R}, and further developed  in  \cite{HK} in the context of nonlinear realization of the ${\cal N}=1$ Maxwell superfield and the component structure of the supersymmetric nonlinear electrodynamics \cite{Kuz} (see additional  refs. in these papers).

To make a comparison between the VA Lagrangian (\ref{finaction'})
\begin{eqnarray} \label{VAlag}
{\cal L}_{VA}= 
 \frac{1}{a} - \frac{i}{4}\bar\psi^{,m}\gamma_{m}\psi
 - \frac{a}{32}[(\bar\psi^{,m}\gamma_{m}\psi)^2{} -
(\bar\psi^{,n}\gamma_{m}\psi)(\bar\psi^{,m}\gamma_{n}\psi)
 ] +  a^2 T^{(3)} 
\end{eqnarray} 
 and the KS Lagrangian  (\ref{KS}) clearer, we present the latter in the  bispinor Majorana representation omitting the terms which have the form of total derivative 
\begin{eqnarray} \label{KSlag} 
{\cal L}_{KS}=
 \frac{1}{a} - \frac{i}{4}{\bar g}^{,m}\gamma_{m} g
- \frac{a}{16}[({\bar g}^{,m}g)^{2} + ({\bar g}^{,m}\gamma_{5}g)^{2}]  \nonumber \\
- (\frac{a}{16})^3[({\bar g}g)^{2} + ({\bar g}\gamma_{5}g)^{2}]
[(\partial^{2}({\bar g}g))^2 +((\partial^{2}({\bar g}\gamma_{5}g))^{2}],
\end{eqnarray}
where $g:= \sqrt{2}G$ and $a:=-1/f^2$, and the relations \cite{Z_Gra} connecting bilinear covariants in the Weyl and the Majorana representations were used. To eliminate the 4-derivative term from ${\cal L}_{KS}$, the expression for the Majorana spinor field $g_a$ in terms of $\psi_a$ has to include its higher derivatives. So, we shall seek for it in the form of a polynomial in the interaction constant $a$
\begin{eqnarray} \label{redef}
g= \psi + a\chi +  a^2\chi_{2} + a^3\chi_{3}, 
\end{eqnarray}
where the sought-for Grassmannian  spinors  $\chi, \, \chi_{2} , \, \chi_{3} $ depend only on $\psi, \, \bar\psi$ and their derivatives. The substitution of the expansion (\ref{redef}) in the KS Lagrangian (\ref{KSlag}) and putting it equal to the VA Lagrangian  (\ref{VAlag}) will produce the equations
 defining the spinors 
 $\chi, \, \chi_{2}$ and  $\chi_{3}$. Thus, the proof of the equivalency of the Lagrangians is reduced to the solutions of these equations.

 The comparison of the terms, having the same degree with respect to the constant $a$ in the redefined KS and the original VA Lagrangians, provides an algorithmic  way to generate the equations under question. By this way we observe that the spinors $\chi_{2}$ and  $\chi_{3}$ don't contribute in the terms linear in $a$ in the redefined $L_{KS}$.
 Thus, it is easy to obtain equation defining the spinor $\chi$.
 Actually,
 the substitution of  (\ref{redef}) into (\ref{KSlag}) and omitting the total derivative term redefines the kinetic term to the form
\begin{eqnarray} \label{kinet}
 - \frac{i}{4}{\bar g}^{,m}\gamma_{m} g =  - \frac{i}{4}\bar\psi^{,m}\gamma_{m}\psi - \frac{i}{2}a(\bar\psi^{,m}\gamma_{m}\chi) +  {\cal O}(a^2).
\end{eqnarray}
The next relevant term from $L_{KS}$  (\ref{KSlag}) is the term linear in $a$ and quartic in the field number. Summing up of the mentioned terms results in the redefined  KS Lagrangian in the linear order in $a$ 
\begin{eqnarray} \label{KSlagred}
{\cal L}_{KS}= \frac{1}{a} - \frac{i}{4}\bar\psi^{,m}\gamma_{m}\psi - \frac{i}{2}a(\bar\psi^{,m}\gamma_{m}\chi) \nonumber \\
- \frac{a}{16}[({\bar\psi}^{,m}\psi)^{2} + ({\bar\psi}^{,m}\gamma_{5}\psi)^{2}] +  {\cal O}(a^2).
\end{eqnarray}
Matching the Lagrangians (\ref{KSlagred}) and (\ref{VAlag}) yields the sought-for equation for $\chi$
\begin{eqnarray}\label{eqn1}
i(\bar\psi^{,m}\gamma_{m}\chi) = 
- \frac{1}{8}[({\bar\psi}^{,m}\psi)^{2} + ({\bar\psi}^{,m}\gamma_{5}\psi)^{2}] +
\nonumber  \\
 \frac{1}{16}[(\bar\psi^{,m}\gamma_{m}\psi)^2{} -
(\bar\psi^{,n}\gamma_{m}\psi)(\bar\psi^{,m}\gamma_{n}\psi)].
\end{eqnarray}

To solve Eq. (\ref{eqn1}) we observe that its terms  have the multiplier $ \bar\psi^{,m}$ which can be canceled resulting in
\begin{eqnarray}\label{eqn2}
\gamma_{m}\chi=  \frac{i}{8}[\psi({\bar\psi}_{,m}\psi) + \gamma_{5}\psi({\bar\psi}_{,m}\gamma_{5}\psi)] -
\nonumber  \\
 \frac{i}{16}[\gamma_{m}\psi (\bar\psi^{,n}\gamma_{n}\psi) -\gamma_{n}\psi
(\bar\psi^{,n}\gamma_{m}\psi)].
\end{eqnarray}
 Multiplication of Eq. (\ref{eqn2}) by  $\gamma^{m}$ results in the general  solution
\begin{eqnarray}\label{solut}
\chi= - \frac{i}{32}[(\gamma_{m}\psi)({\bar\psi}^{,m}\psi) + (\gamma_{m}\gamma_{5}\psi)({\bar\psi}^{,m}\gamma_{5}\psi)] -
\nonumber  \\
 \frac{i}{64}[3\psi (\bar\psi^{,m}\gamma_{m}\psi) + (\Sigma_{mn}\psi)
(\bar\psi^{,n}\gamma^{m}\psi)].
\end{eqnarray}
 
 Substitution of (\ref{solut}) in (\ref{redef}) yields the explicit expression connecting the KS and the VA realizations of the NG field up to terms linear in 
 $a$
\begin{eqnarray} \label{connect}
\sqrt{2}G= \psi[1 +  \frac{3ia}{64}(\bar\psi^{,m}\gamma_{m}\psi)]  -
 \frac{ia}{32}[(\gamma_{m}\psi)({\bar\psi}^{,m}\psi) + 
\nonumber  \\
(\gamma_{m}\gamma_{5}\psi)({\bar\psi}^{,m}\gamma_{5}\psi) -
 \frac{1}{2} (\Sigma_{mn}\psi)(\bar\psi^{,n}\gamma^{m}\psi)] + {\cal O}(a^2).
\end{eqnarray}

The quadratic terms in $a$ are restored through the substitution of $\chi$ (\ref{solut}) into the expansion (\ref{redef}),  and subsequently repetition of the above considered procedure with respect to the quadratic terms in $a$.
Having fulfilled this, one can find  $\chi_2$, and after to repeat again similar procedure with respect to the cubic terms in the constant $a$.
 As a result, one can obtain the explicit expression for the KS field $G$ through the VA field $\psi$ and to conclude about the expected equivalency between the KS and VA  Lagrangians.

\section{Discussions}

Here we presented an independent  algebraic 
 proof of the cancellation of 4-derivative terms in 
the $D=4 \  {\cal N}=1$ VA action using the 
Majorana bispinor representation and the Fierz rearrangements. 
The Majorana representation may simplify the  investigation 
of such cancellations in the case of extended 
supersymmetries and/or of the higher dimensional spaces.
We observed that the cancellation results in the difference 
between the Komargodski-Seiberg superfield \cite{Seib} and the 
Volkov-Akulov \cite{VA} actions.

The difference gives rise to the question whether the  KS Lagrangian is  
  equivalent to the VA Lagrangian?
The second question arising  from the cancellation concerns 
its presence in the NG fermion interactions with other fields. 
We found out that the cancellation occurs in the coupling
of the fermionic NG field  with massive Dirac fields. 
It yields the maximal number of the NG fermions $n_{NGf}$ and their 
derivatives $n_{NGd}$ in the interaction Lagrangian which equals six 
and three, respectively. The maximal numbers $n_{NGf}=6$
and  $n_{NGd}=3$ are the same as in the VA action describing the NG fermion 
interactions between themselves.
The observation poses the issue of restoration of superfield  
Lagrangian of interactions which uses realization of the NG fermionic field coinciding with the one in the VA Lagrangian with couplings.  
A way to solve these issues  implies the construction of the explicit  expression connecting the KS and the VA  realizations  of the NG field. The representation of the KS  fermion field through the VA  field has to  contain terms  with its  derivatives. We discussed the problem and found the  explicit formula connecting the VA and the  KS realizations of NG field up to the first order in the interaction constant $a$.  The substitution of the expression into the  KS action reduced it  to the VA action. It points to the expected equivalence of these actions in all orders in  $a$.  The equivalency problem posed in \cite{Zhep},   has recently been discussed in \cite{L3W} with pointing to some difficulties appearing on the way.

 Taking into account the recent application of the formalism of $ {\cal N}=1$  constrained  superfields in the minimal supersymmetric  standard model (MSSM), as well as its generalizations to $ {\cal N}$-extended supersymmetric models  (see e.g. \cite{ADGT}, \cite{AB}), it is interesting to study the above considered  kind of  cancellations in these models. Availability of an explicit formula connecting VA and KS realizations of NG field could  simplify the phenomenological analysis of the mentioned and other new models.

\noindent{\bf Acknowledgments}

I would like to thank Paolo Di Vecchia and Fawad Hassan for the 
interesting discussions and critical remarks
and Sergei Kuzenko for the letter concerning the paper \cite{Kuz}.
Also, I am owed to Eugenii Ivanov, Sergei Ketov, Zohar Komargodski and Sergei Kuzenko for their helpful comments concerning  \cite{Zhep} and new references added.
I am grateful to the Department of Physics of Stockholm University 
and Nordic Institute for Theoretical Physics Nordita for kind hospitality. 
This research was supported in part by Nordita.

\end{document}